\documentclass[aps,prl,twocolumn,superscriptaddress,showpacs,showkeys]{revtex4-1}
\usepackage{graphicx}
\usepackage{bm}

\begin{document}

\title{All-optical magnetization switching by two-frequency pulses using the plasmon-induced inverse Faraday effect in a magneto-plasmonic structure}
\author{Song-Jin Im}
\email{ryongnam31@yahoo.com}
\affiliation{Department of Physics, Kim Il Sung University, Taesong District, 02-381-4410 Pyongyang, Democratic People's Republic of Korea}
\author{Ji-Song Pae}
\affiliation{Department of Physics, Kim Il Sung University, Taesong District, 02-381-4410 Pyongyang, Democratic People's Republic of Korea}
\author{Chol-Song Ri}
\affiliation{Department of Physics, Kim Il Sung University, Taesong District,  02-381-4410 Pyongyang, Democratic People's Republic of Korea}
\author{Kum-Song Ho}
\affiliation{Department of Physics, Kim Il Sung University, Taesong District,  02-381-4410 Pyongyang, Democratic People's Republic of Korea}
\author{Joachim Herrmann}
\email{jherrman@mbi-berlin.de}
\affiliation{Max-Born-Institute for Nonlinear Optics and Short Pulse Spectroscopy, Max-Born-Str. 2a,
D-12489 Berlin, Germany}
\date{\today}
\begin{abstract}
In this Letter we study the generation of quasi-static magnetic fields by the plasmon-induced inverse Faraday effect and propose a magneto-optical waveguide structure for achieving magnetization switching at sub-ps time in a nano-confined magneto-optical structure. In particular we show that the direction of the generated quasi-static field in a magneto-optical dielectric cavity side-coupled to a metal-insulator-metal (MIM) waveguide depends sensitively on the wavelength of the surface plasmon polaritions (SPP). This phenomenon could open up a new energy-efficient ultrafast method for nano-confined all-optical magnetization switching by two-frequency pulses.
\end{abstract}

\maketitle

In a magneto-optical material irradiated by circularly polarized light an effective quasi-static magnetic field along the wave vector is induced. This phenomenon called the inverse Faraday effect (IFE) \cite {Pitaevskii1961,VanderZiel1965} is a counterpart of the Faraday effect which rotates the linear polarization of light propagating through a magneto-optical material in the presence of a magnetic field. Left- and right-handed circular polarized light induce the magnetization with opposite direction and the magnetization vanishes for linearly polarized waves. Several theoretical approaches have been developed to provide physical insight on the effect (see e.g. \cite{Kurkin2008,Popova2012,Qaiumzadeh2013,Battiato2014,Berrita2016}).

Recently, the inverse Faraday effect has attracted much attention because of its potential impact for ultrafast all-optical switching of magnetization in thin magnetic films induced by ultrafast pulses \cite{Beaurepaire1996,Stanciu2007,Kirilyuk2010} which opens the possibility for magnetic data storage with unprecedented speed. The influence of the helicity of the circular polarized laser pulses in magnetization switching has been attributed to the IFF but in later studies depending on the magnetic material other explanations has been proposed which remain up to now in investigation. Recent studied ferromagnetic thin films indicate the basic role of the IFE in helicity-depending magnetization in these materials \cite{Cornelissen2016,John2017}. However, the spot size of the input pulses limit the density memory modules and sets limits for the lowest requested pulse energy. Moreover, all-optical magnetic switching by free space circularly polarized light is not fully waveguide-integrated and not compatible with highly integrated photonic circuits.

In order to improve the recording density a method is requested which allow sub-wavelength spatial resolution of light pulses for magnetic recording. 
The IFE in plasmonic structures illuminated by free space circularly polarized light can be significantly amplified by the plasmon field enhancement caused by localized surface plasmons \cite {Belotelov2010,Hamidi2015,Korff2015,Dutta2017}. Besides laser pulses has been focused by using plasmonic nanoantena structures to achieve nanoscale confinement of all-optical magnetization switching \cite{Stipe2010,Liu2015,Guyader2015}.

In this paper, we study a modified version of nanoscale IFE in which instead of free space circular polarized optical pulses free running surface plasmon polaritons (SPPs) induce a magnetic field. SPPs exhibit a longitudinal component of the electric field, therefore the vector product of E and its complex conjugate E* do not vanishes. This means that even for linearly polarized input beams a magnetization can be induced by running SPPs generated by the incident p-polarized light pulses, but the plasmons are not circularly polarized in the customary sense. Recently, based on this phenomenon an IFE-related third-order nonlinearity of surface plasmon polaritons was predicted \cite {Im2017_1}. By using the plasmon-induced IFE we propose and study a magneto-optical structure consisting of a magneto-optical cavity side-coupled to a metal-insulator-metal (MIM) waveguide and demonstrate the possibility for all-optical magnetic switching by two-frequency sub-ps pulses.

The inverse Faraday effect (IFE) can be  described for a lossless opto-magnetic material by the thermodynamical potential $\Phi  = {\varepsilon _0}{\beta _{ijk}}E{}_iE{}_j{M_k}$  where $\vec{E}$ is the electric field of the light with frequency $\omega$. Therefore for isotropic materials we find
\begin{eqnarray}
{\vec H_{{\rm{eff}}}} =  - i{\varepsilon _0}\beta \vec E \times {\vec E^ * },
\label{eq1}
\end{eqnarray}    
where $\beta$  is the magneto-optical susceptibility. To achieve a high magnitude of the effective magnetic field ${\vec H_{{\rm{eff}}}}$, a material with large magneto-optical susceptibility $\beta$  as well as a value of $\vec m =  - i\vec E \times {\vec E^ * }$  inside the material 
as large as possible are requested.
 For linearly polarized light, $\left| {\vec m} \right| = 0$.  For circularly polarized light propagating along the $\hat x$ -axis we have $\vec m = (m,0,0)$ with  
         
\begin{eqnarray}
m =  \pm {\left| {\vec E} \right|^2},
\label{eq2}
\end{eqnarray}  
where the sign of $m$ is dependent on the helicity of the circularly polarized light.

\begin{figure}
\includegraphics[width=0.3\textwidth]{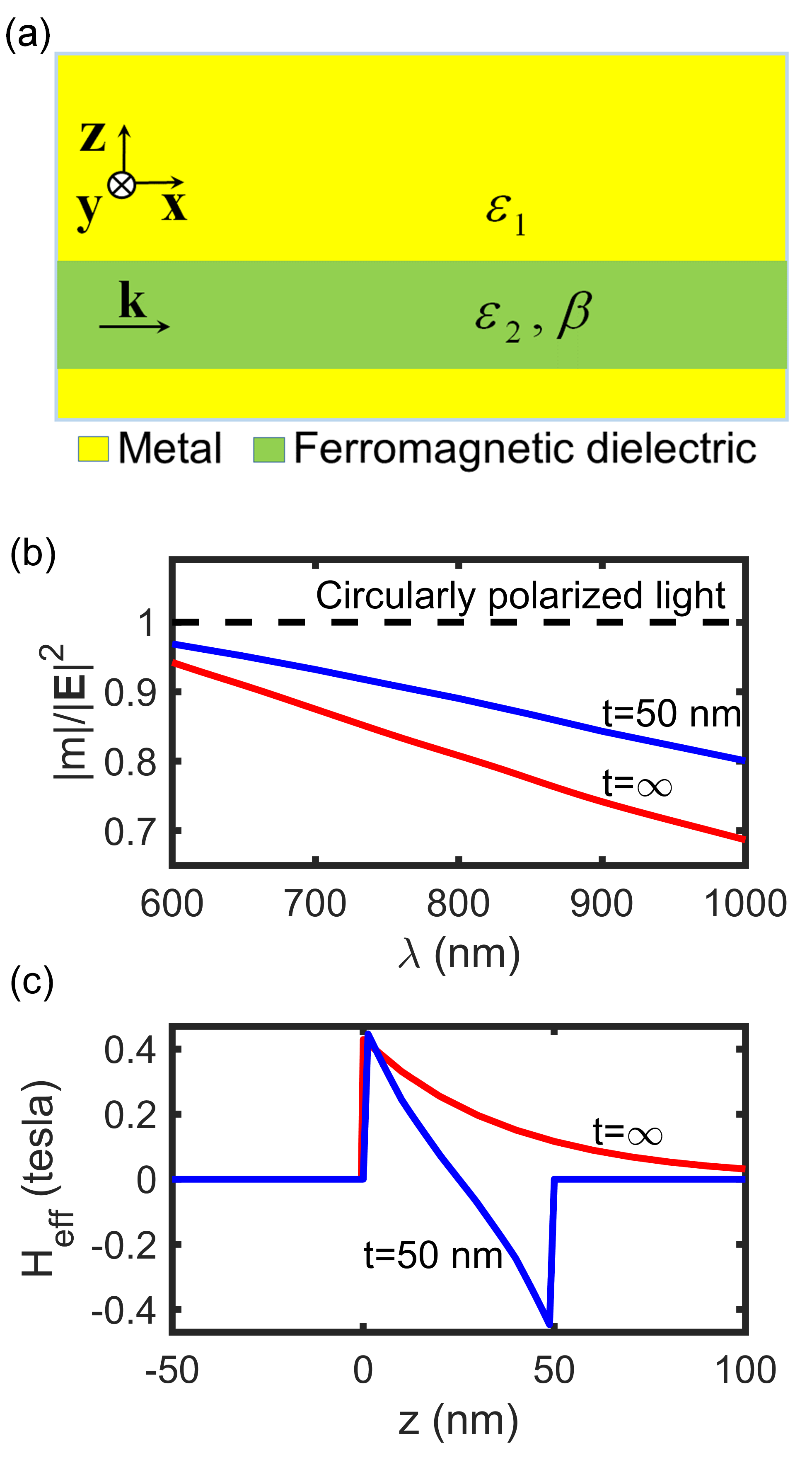}
\caption{(a) The geometry of metal-insulator-metal (MIM) waveguide with an insulator layer composed from a ferromagnetic dielectric. (b) Wavelength dependence of the value of $\left| m \right|/{\left| {\vec E} \right|^2}$ in the MIM waveguide. (c) Transverse distribution of the effective magnetic field $H_{{\rm{eff}}}$ in the MIM waveguide at the wavelength $\lambda=800$ nm. The experimental data of the permittivity of silver \cite {Johnson1972} have been used as the metal permittivity ${\varepsilon _1}$. The experimental data of the permittivity and and the magneto-optical susceptibility of  Bi-substituted iron garnet (BIG) \cite {Dutta2017} have been used as ${\varepsilon _2}$ and $\beta$, respectively. The blue curve is for the case of the waveguide thickness $t=50$ nm and the red curve for the case of a single metal-dielectric interface, $t=\infty$.  The mode power of the incident SPPs has been assumed to be 10 W/$\mu$m.}
\label{fig:1}
\end{figure}     

 \begin{figure}
	\includegraphics[width=0.3\textwidth]{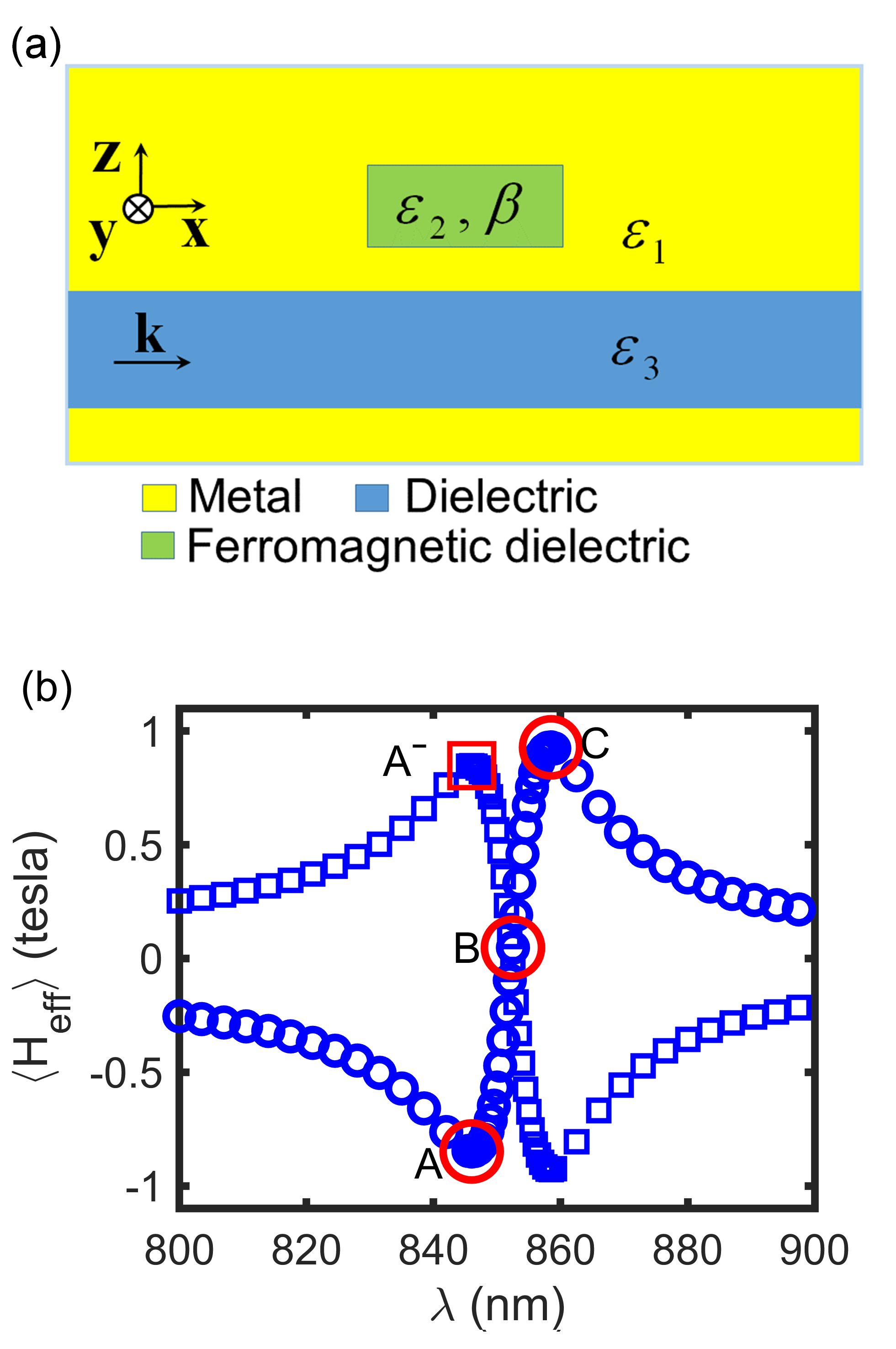}
	\caption{(a) The geometry of the magneto-optical cavity side-coupled to a MIM waveguide. (b) The average effective magnetic field $\left\langle H_{{\rm{eff}}} \right\rangle$ versus the wavelength. $\left\langle \right\rangle$ denotes the average value over the magneto-optical cavity (the green part of (a)).  The blue circles correspond to the case of SPPs incident from the left port of the MIM waveguide and the blue squares to the case of SPPs incident from the right port. For the MIM waveguide (the blue part of (a)), ${\varepsilon _{3}} = 1$ and a thickness of 30 nm have been assumed. The thickness of the magneto-optical cavity (the green part of (a)) is 20 nm and the length is 40 nm. The permittivity and the magneto-optical susceptibility of the ferromagnetic dielectric and the metal permittivity are the same as in Fig.~1.}
	\label{fig:2}
\end{figure}           

\begin{figure*}
	\includegraphics[width=0.8\textwidth]{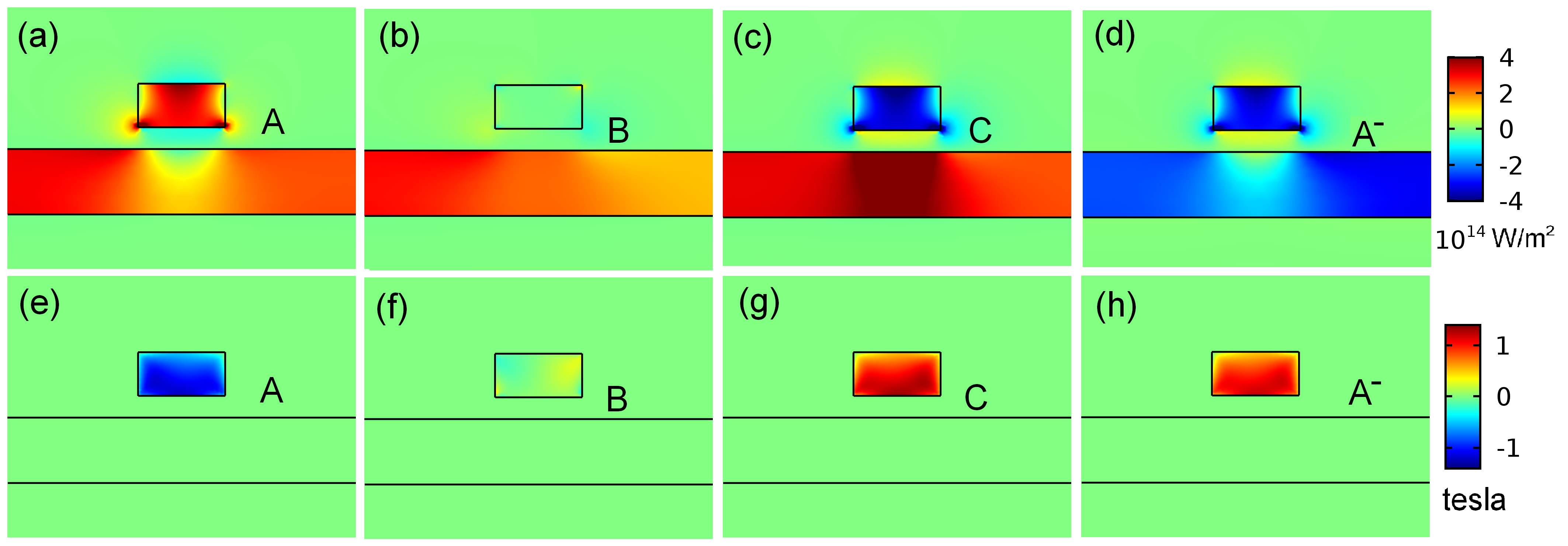}
	\caption{Near field distributions of the x-component of the pointing vector ((a), (b), (c) and (d)) and of the effective magnetic field $H_{{\rm{eff}}}$ ((e), (f), (g) and (h)) at the points of A, B, C and A$^{-}$ of Fig. 2(b).}
	\label{fig:3}
\end{figure*}           

\begin{figure}
	\includegraphics[width=0.3\textwidth]{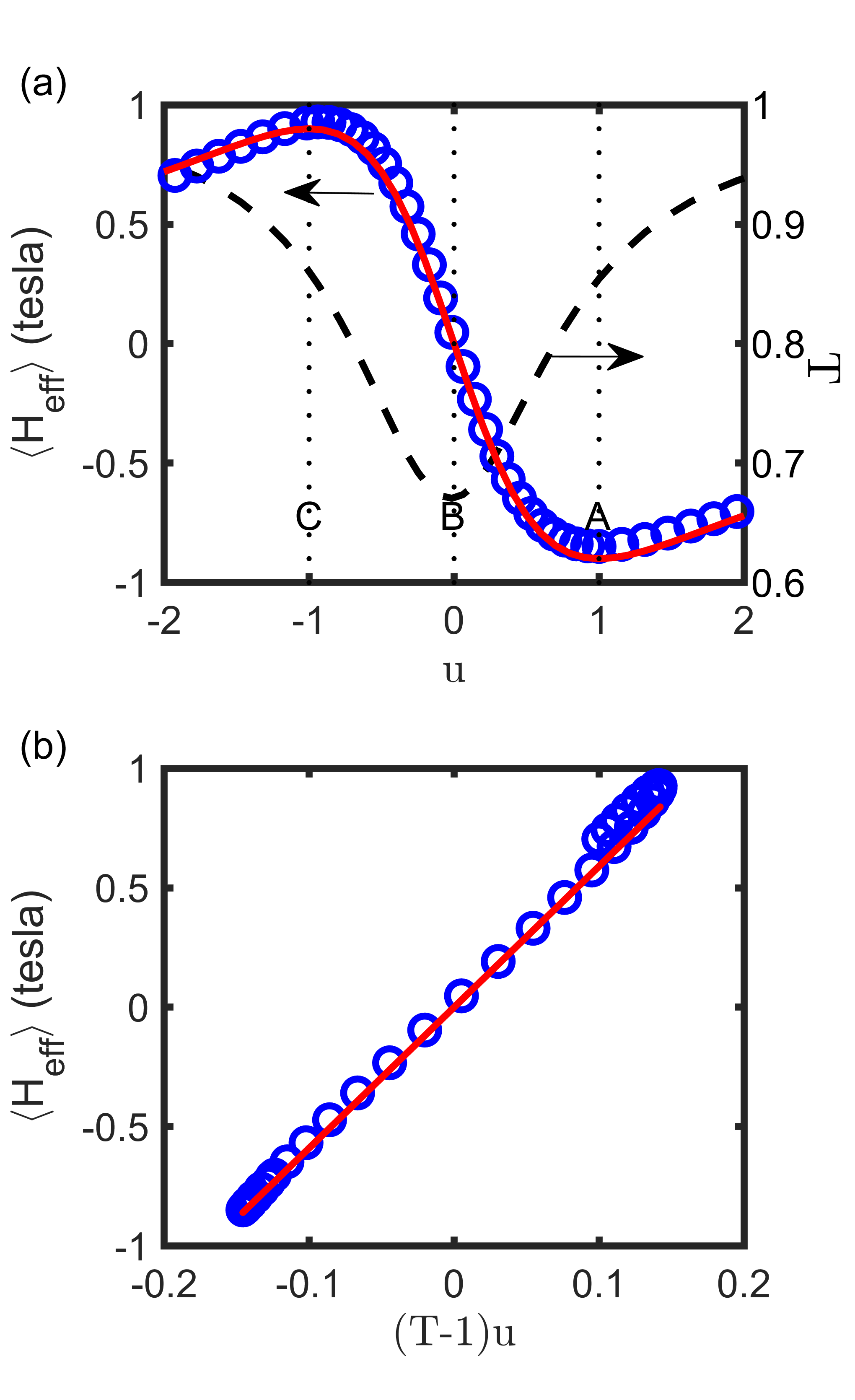}
	\caption{(a) Average effective magnetic field $\left\langle H_{{\rm{eff}}} \right\rangle$ versus the normalized detuning $u = (\omega  - {\omega _0})/\gamma$, where $\gamma$ has been calculated as the half width of the resonance from the numerical results of transmission of the SPPs through the MIM waveguide (the black dashed curve). (b) The linear relationship between the $(T - 1)u$ and $\left\langle H_{{\rm{eff}}} \right\rangle$. The blue circles and the red curves are obtained from the numerical solution of Maxwell's equations  and from Eq. (5), respectively. Other parameters are the same as in Fig. 2.}
	\label{fig:4}
\end{figure}                 

Light cannot penetrate into a thin metallic layer, but under appropriate conditions (as e.g. by using the Kretschman configuration for p-polarized light or a grating structure) surface plasmon-polaritons (SPP) can be excited moving along the surface. SPPs exhibit a longitudinal component of the electric field, therefore the polarization-depending value of $m$ do not vanishes. This means that even for linearly polarized input beams a magnetization can be induced by plasmons, but the plasmons are not circularly polarized in the customary sense. 
Let us calculate the quasi-static magnetic field induced by the longitudinal polarized SPP. 
According to Eq. (1) the magnetic field is determined by the entity $\vec m = (0,m,0)$ where $m$ is defined by 
\begin{eqnarray}
m =  - i(E_x^ * {E_z} - {E_x}E_z^ * ).
\label{eq3}
\end{eqnarray}                
Here the wavevector is along the $\hat x$-axis and the electric field vector is in the $x$-$z$  plane. If we substitute the expression for the electric field distribution of a single metal-dielectric interface \cite {Maier2007} in Eq. (3) we find
\begin{eqnarray}
\left| {{m}} \right| = 2{\mathop{\rm Re}\nolimits} (\sqrt { - {\varepsilon _2}{\varepsilon _1}} )/({\varepsilon _2} - \left| {{\varepsilon _1}} \right|) \cdot {\left| {\vec E} \right|^2},
\label{eq4}
\end{eqnarray}
where ${\varepsilon _2}$ and ${\varepsilon _1}$   are the permittivities of the dielectric and the metal, respectively.  For a dielectric  ${\rm{sign}}({m})$=1 and For a metal ${\rm{sign}}({m})$ =-1. The value of $\left| {\vec m} \right|/{\left| {\vec E} \right|^2}$ in a single metal-dielectric interface in dependence on the wavelength is shown in the red curve of Fig. 1(b). Near the plasmon resonance ${\mathop{\rm Re}\nolimits} ({\varepsilon _2} + {\varepsilon _1}) = 0$, $\left| {m} \right|/{\left| {\vec E} \right|^2} \approx 1$  which is close to the case of excitation by free space circularly polarized light.  Away from the resonance, this parameter is reduced, but still maintaining a moderate value of 0.5. 
In Fig. 1(b) and (c) a magneto-optical metal-insulator-metal (MIM) waveguide as shown in Fig. 1(a) is studied with a dielectric layer composed of Bi-substituted iron garnet (BIG). 
In Fig. 1(b) the value of $\left| {\vec m} \right|/{\left| {\vec E} \right|^2}$ in dependence on the wavelength is shown. As seen this parameter increases with the reduction of the thickness of the dielectric layer. In Fig. 1(c) the transverse distribution of the IFE-induced quasi-static magnetic field is presented where the induced magnetic field is 0.4 tesla for an incident mode power of 10 W/$\mu$m and its direction on the downside interface is opposite to that in the upside interface.

Based on the plasmon-induced IFE we now study magnetization switching by a magneto-optical cavity side-coupled to a MIM waveguide as shown in Fig. 2(a). 
The average effective magnetic field $\left\langle H_{{\rm{eff}}} \right\rangle$ denotes the average value over the magneto-optical cavity (the green part of Fig. 2(a)). $\left\langle H_{{\rm{eff}}} \right\rangle$ is sensitive to the wavelength near the resonance and perceives a sign change (the points A and C of Fig. 2(b)). It becomes zero at the exact resonance (the point B of Fig. 2(b)). The sign of $\left\langle H_{{\rm{eff}}} \right\rangle$ is reversed for backward propagating SPPs (the point A$^{-}$ of Fig. 2(b)). 

Fig. 3 shows the near field distribution of the x-component of pointing vector ((a), (b), (c) and (d)) and $H_{{\rm{eff}}}$ ((e), (f), (g) and (h)) for the points A, B, C and A$^{-}$ of Fig. 2(b) calculated by the numerical
solution of Maxwell's equations. At the point A, the effective magnetic field $H_{{\rm{eff}}}$ has a negative value of 1 tesla inside the magneto-optical cavity for SPPs moving into the direction along the x-axis. At the points C and A$^{-}$, the effective magnetic field has the same value of 1 tesla, but the energy flow is into the reverse direction. At the point B, both the effective magnetic field and the energy flow disappear because of the perfect standing wave mode incide the cavity at the exact resonance.

The resonance of $\left\langle H_{{\rm{eff}}} \right\rangle$ has a Lorentzian-like profile which can be approximated by

\begin{eqnarray}
\nonumber
\left\langle H_{{\rm{eff}}} \right\rangle =  - 2au/(1 + {u^2}) ,\\
u = (\omega  - {\omega _0})/\gamma  .
\label{eq5}
\end{eqnarray}
Here $\gamma$ is the decay rate, ${\omega _0}$ the resonance frequency and $a$ a coefficient. On the other hand, the transmission $T$ through the magneto-optical structure is described by $(1 - T) = (1 - {T_0})/(1 + {u^2})$ \cite {Im2016}. The results by Eq. (5) (the red curve of Fig. 4(a)) well agree with the results of $\left\langle H_{{\rm{eff}}} \right\rangle$ (the blue circles of Fig. 4(a)) obtained by the numerical solution of Maxwell's equations. $\gamma$ has been calculated from the numerical results for the transmission (the black dashed curve). As shown in Fig. 4(a), $\left\langle H_{{\rm{eff}}} \right\rangle$ has a positive maximum value at $u =  - 1$, a negative minimum value at $u= 1$ and is zero at $u= 0$. The linear relationship between the value of $(T - 1)u$ and $\left\langle H_{{\rm{eff}}} \right\rangle$ is clearly demonstrated as shown in Fig. 4(b) by the numerical results of the solutions of Maxwell's equations. This suggests that the plasmon-induced magnetic field in the magneto-optical cavity can be indirectly measured by measuring the transmission of the waveguide plasmon.

Let us estimate the order of energy consumption needed for reversable binary magnetic recording. We note that magnetization of BIG saturates for 0.15 tesla (see the Methods of Ref. \cite{Davoyan2014}). From Fig. 2 and 3 one predicts that a power of 1.5 W/$\mu$m of the incident pulses can produce an effective magnetic field on the order of 0.15 tesla. The needed energy is roughly estimated as $1 {\rm{W}}/{\mu}{\rm{m}} \cdot w \cdot \tau \cdot$, where $w$ is the lateral size of the waveguide in the y-axis direction and $\tau$ the full-width at half maximum (FWHM) of the incident pulse. 
We take $w \simeq$ 1 $\mu$m and $\tau  \simeq$ 100 fs under the assumption that a lateral size on the order of 1 $\mu$m does not induce noticeable deterioration to the device performance \cite{Cai2009} and the magnetization switching can be achieved on a time scale of 100 fs \cite {Vahaplar2012}. Thus the energy consumption is on the order of 100 fJ/bit. We remind that a laser pulse fluence on the order of 1 mJ/cm$^2$ \cite{Dutta2017} is needed to induce an effective magnetic field on the order of 0.1 tesla inside a ferromagnetic film illuminated by circularly polarized light beam with a beam diameter of tens of $\mu$m, thus the needed energy for the traditional IFE is on the order of 10 nJ/bit. This rough estimation shows that the IFE induced by quasi-resonance in a magneto-optical cavity side-coupled to a MIM waveguide requires several orders of magnitude smaller pulse energy for the generation of a magnetic field of 0.1 tesla compared with the traditional IFE. This significant reduction is attributed to the strong confinement in the MIM waveguide and the cavity modes.

In conclusion, we investigated a modified version of the inverse Faraday effect induced by running SPPs with a longitudinal component of the electric field vector in magneto-optical waveguides. We calculated the effective magnetic field in a metal-insulator-metal (MIM) waveguide with a dielectric magneto-optical layer composed of Bi-substituted iron garnet (BIG) and found that its direction on the upside interface is opposite to that of the downside interface. Based on the SPP-induced IFE we investigated magnetization switching by using a magneto-optical cavity side-coupled to a MIM waveguide. The SPP-induced magnetization inside the magneto-optical cavity has a Lorentzian-like resonance profile. Magnetization reversal can be achieved by using two-frequency pulses with opposite frequency detuning from the resonance so that the detuning parameter $u$  has a value 1 or -1. This approach for nano-scale magnetization switching is energy-efficient, fully waveguide-integrated, compatible with integrated nano-photonic circuits and could be promising for all-optical magnetic recording in highly integrated photonic circuits.

\bibliography{pife}
\end{document}